\documentclass[useAMS,usenatbib]{mn2e}
\pdfoutput=1
\usepackage{graphicx}
\usepackage{dcolumn}
\usepackage{bm}
\usepackage{amssymb,amsmath,bm}
\usepackage{color}
\usepackage[colorlinks,linkcolor=red,citecolor=blue,urlcolor=blue]{hyperref}
\usepackage{multirow}
\usepackage[utf8]{inputenc}
\usepackage{balance}
\usepackage{enumitem}
\usepackage{lipsum}
\usepackage{ulem}

\newcommand{\aap}{A\&A}

\newcommand{\apjs}{ApJS}

\newcommand{\apjl}{ApJL}

\title[Simulation on Particle Acceleration in a Nonrelativistic Quasiparallel shock]{Kinetic Simulation on Electron, Proton and Helium Acceleration in a Nonrelativistic Quasiparallel shock}
\author[Jun Fang et al.]{Jun Fang$^{1,2}$\thanks{E-mail: fangjun@ynu.edu.cn}, Qi Xia$^1$, Shiting Tian$^1$, Liancheng, Zhou$^1$, Huan Yu$^3$ \\
                      $^{1}$Department of Astronomy, Yunnan University, Kunming 650091, China\\
                      $^{2}$Key Laboratory of Astroparticle Physics of Yunnan Province, Yunnan University, Kunming 650091, China\\
                      $^{3}$Department of Physical Science and Technology, Kunming University, Kunming 650214, China
                      }

\begin{document}
  \date{\today}
  \pagerange{1--6} \pubyear{2020}
  \maketitle

\begin{abstract}
  In addition to electrons and protons, nonrelativistic quasiparallel shocks are expected to possess the ability to accelerate heavy ions. The shocks in supernova remnants are generally supposed to be accelerators of the Galactic cosmic rays, which consist of many species of particles.  We investigate diffusive shock acceleration (DSA) of electrons, protons and helium ions in a nonrelativistic quasiparallel shock through 1D particle-in-cell (PIC) simulation with a helium-to-proton number density ratio of  $0.1$, which is relevant for the Galactic cosmic rays. The simulation indicates that waves can be excited by the flow of the energetic protons and helium ions upstream of the nonrelativistic quasiparallel shock  with a sonic Mach number of 14 and an alfv\'{e}n Mach number of 19.5 in the shock rest frame, and the charged particles are scattered by the self-generated waves and accelerated gradually. Moreover, the spectra of the charged particles downstream of the shock are thermal plus a nonthermal tail, and the acceleration is efficient  with about $7\%$ and $5.4\%$ of the bulk kinetic energy transferred into the nonthermal protons and helium ions in the near downstream region at the end of the simulation, respectively.
\end{abstract}


\section{Introduction}\label{sec:intro}
\begin{figure*}
        \centering
        \includegraphics[width=0.33\textwidth]{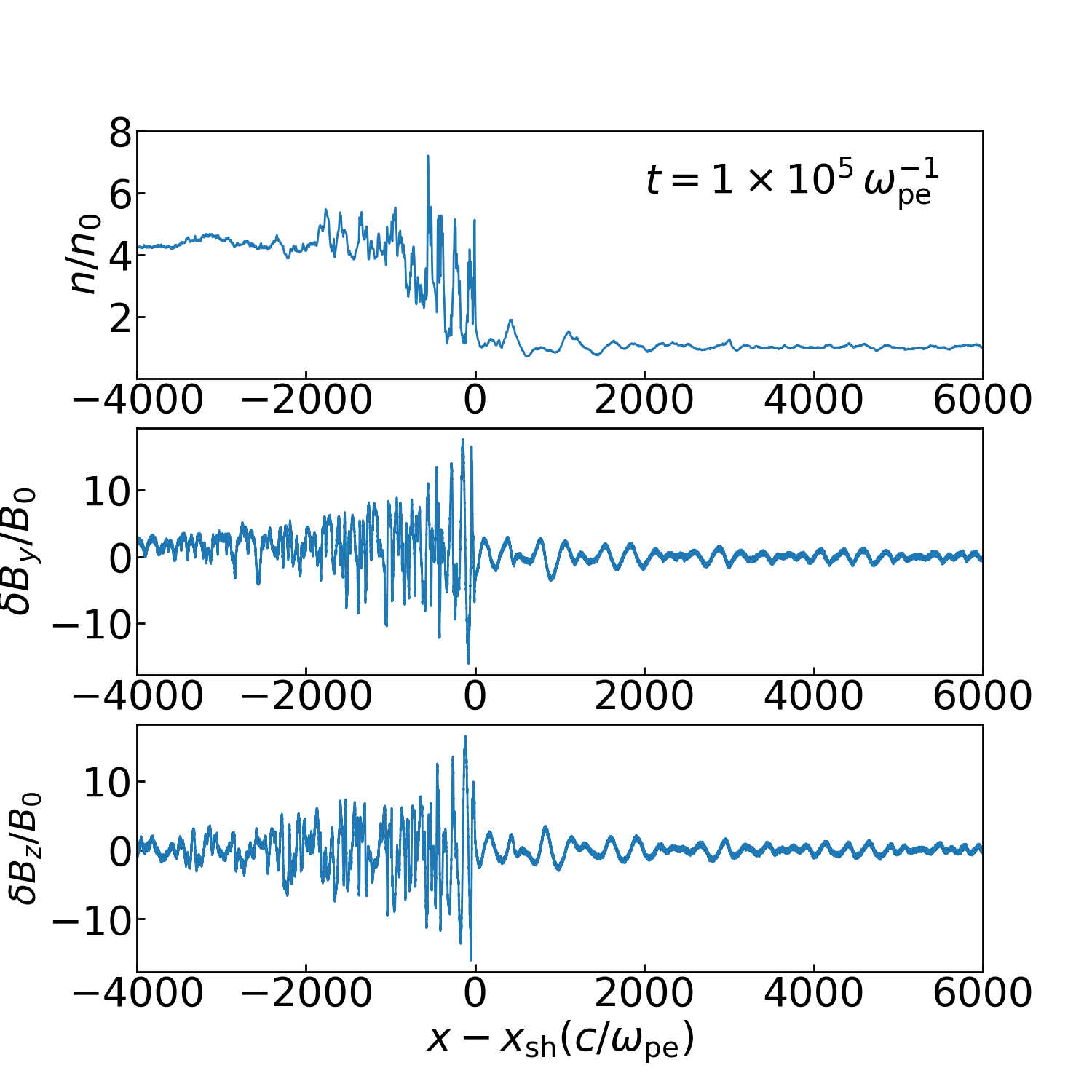}
        \includegraphics[width=0.33\textwidth]{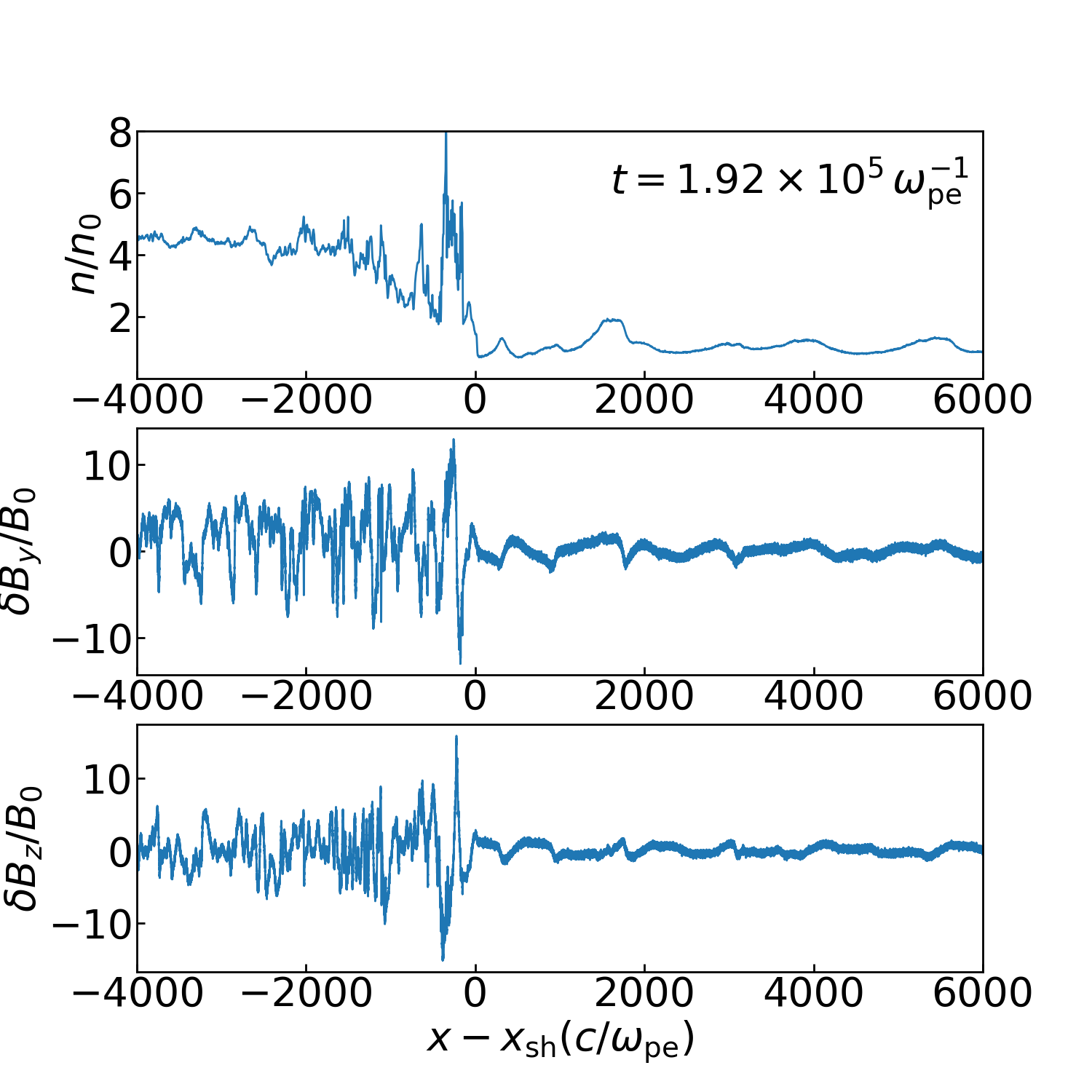}
        \includegraphics[width=0.33\textwidth]{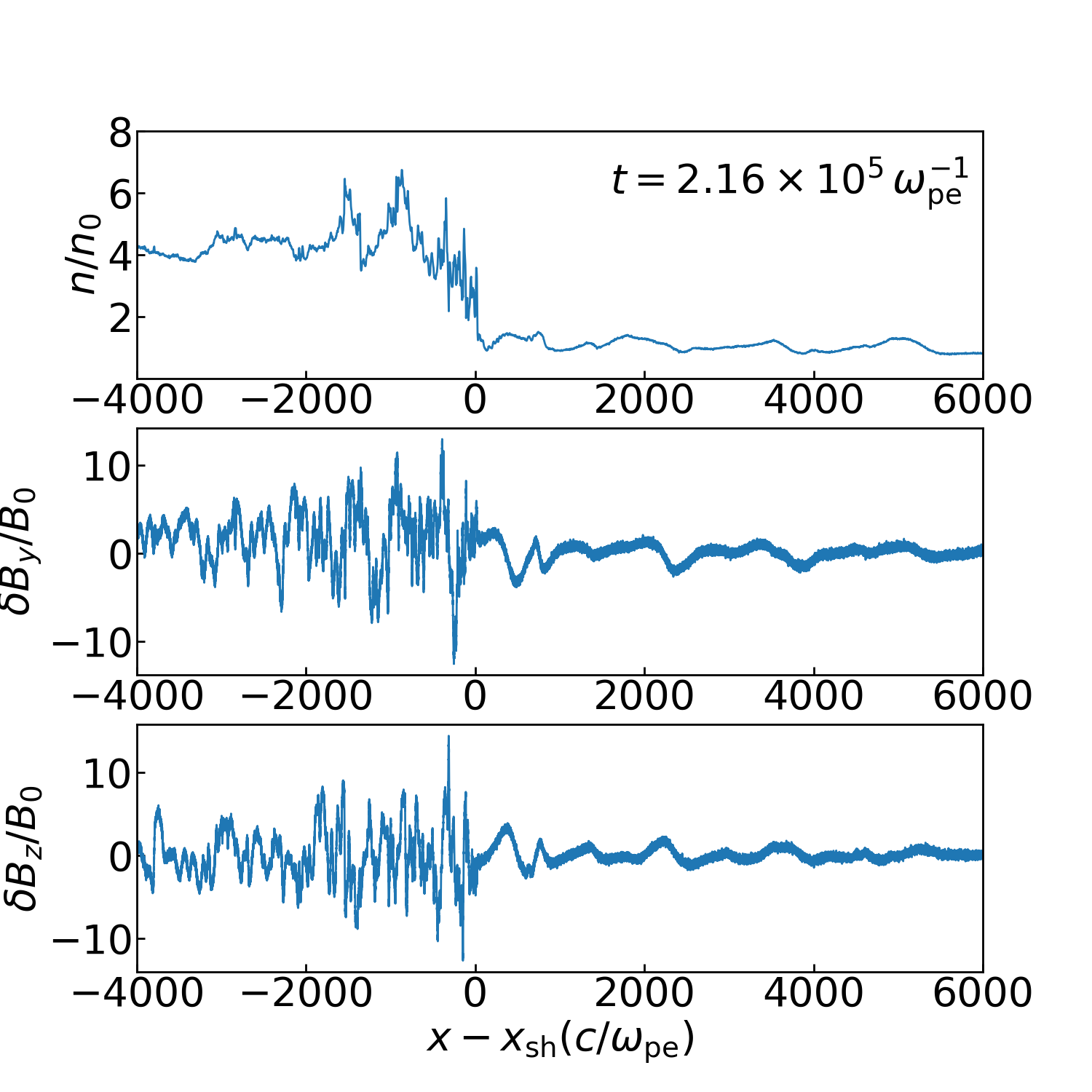}
        \caption{Profile of proton number density normalized by $n_0$ (top panels), the self-generated magnetic field $\delta B_y$ (middle panels) and $\delta B_z$ (bottom panels) for the shock at $t=1\times10^5\,\omega_{\mathrm{pe}}^{-1}$, $1.92\times10^5\,\omega_{\mathrm{pe}}^{-1}$ and $2.16\times10^5\,\omega_{\mathrm{pe}}^{-1}$ (from left to right), respectively. The shock is located at $x_{\mathrm{sh}}=6.25\times 10^3\,c/\omega_{\mathrm{pe}}$ (left panels), $1.18\times10^4\,c/\omega_{\mathrm{pe}}$ (middle panels) and $1.28\times 10^4\,c/\omega_{\mathrm{pe}}$ (right panels) for $t=1\times10^5\,\omega_{\mathrm{pe}}^{-1}$, $1.92\times10^5\,\omega_{\mathrm{pe}}^{-1}$ and $2.16\times10^5\,\omega_{\mathrm{pe}}^{-1}$, respectively, and the $x$ coordinate is shifted by setting  the shock location to zero. }
        \label{fig:shockstruc}
\end{figure*}

Nonrelativistic shocks in supernova remnants (SNRs)  are widely thought as efficient particle accelerators to produce energetic cosmic rays in the Galaxy. A shock in a SNR is induced as the high-speed supernova ejecta expanding into the ambient medium, and a part of the kinetic energy of the ejecta can be transferred to the cosmic rays accelerated by the shock. Multiband observations on some SNRs indicate that electrons and ions can be accelerated to energies above $100$ TeV \citep{2012A&ARv..20...49V}. Nonisotropic explosion of the progenitor of a SNR and the inhomogeneity of the ambient medium usually induces a complex morphology of the remnant. Moreover, as indicated in the radio, X-ray and $\gamma$-ray morphologies of SN1006 \citep{2003ApJ...589..827B,2003ApJ...586.1162L,2004A&A...425..121R,2010A&A...516A..62A,2013AJ....145..104R}, the acceleration efficiency on the accelerated particles in a SNR can vary azimuthally along the border of the remnant. These properties indicated in the multiband observations on SNRs relate to the acceleration mechanisms involved in the shocks.

DSA is widely accepted as the primary process to accelerate ions to relativistic energies for the nonrelativistic shocks in SNRs  \citep{1978ApJ...221L..29B,1978MNRAS.182..147B}. In the process,  the charged particles are scattered by the magnetic turbulence upstream and downstream of the shock, and the particles gain energy cumulatively when they pass through the shock back and forth. Electrons, however, must undergo pre-acceleration to have large enough gyro radius to passthrough the shock with a thickness comparable to the ion gyro radius, and then they can participate in the DSA process to be accelerated \citep{2014ApJ...794..153G}.

Mechanisms of particle acceleration involved in shocks can be investigated using kinetic simulations. Based on 1D particle-in-cell (PIC) simulations, \citet{2015PhRvL.114h5003P} indicated that both protons and electrons could undergo DSA after the preheating via shock drift acceleration (SDA) at nonrelativistic quasiparallel shocks. Moreover, the particles downstream of the shocks have power-law distributions, i.e., the momentum distribution $f(p)\propto p^{-4}$, which is consistent with the theoretical prediction of DSA. For nonrelativistic quasi-perpendicular collisionless shocks, electrons can be preheated by SDA due to the reflection by magnetic mirror near the shock and the scattering off the upstream waves, then they can be injected into DSA to form a power-law downstream spectrum \citep{2020ApJ...897L..41X}. The onset of ions into the DSA process at a quasi-perpendicular nonrelativistic shock was also identified based on 1D PIC simulations \citep{2021ApJ...921L..14K}.

Acceleration of ions at shocks can also be investigated via hybrid PIC simulations. Different from PIC approach, hybrid PIC simulations treat ions kinetically, but electrons are represented by a fluid. Hybrid PIC simulations on particle acceleration at non-relativistic shocks indicate that DSA is efficient to accelerate ions in quasi-parallel shocks \citep{2014ApJ...783...91C}. At oblique shocks, the acceleration of ions is inefficient above an energy of $\sim 5 m v_{\rm sh}^2$, where $m$ is the ion mass, $v_{\rm sh}$ is the upstream fluid velocity in the downstream reference frame \citep{2014ApJ...783...91C}. Furthermore,  hybrid PIC simulations on acceleration of different ions with different mass/charge ratios in nonrelativistic shocks  indicate that incompletely ionized ions can be preferentially accelerated \citep{2017PhRvL.119q1101C}.

Shock acceleration of particles including electrons, protons and heavier ions needs more investigation with fully kinetic PIC simulations.  \citet{2020arXiv200307293S} used 1D PIC simulations to study shock formation and particle acceleration in a quasi-parallel shock imposed by a plasma composed of electrons, protons, helium and carbon ions, and the results indicated that $\sim 12-15\%$ of the kinetic energy of the upstream flow can be can be injected into the accelerated particles. In their simulations, a supra-thermal tail is produced in the spectrum of the each particle component downstream of the shock, but a power-law distribution is not formed due to the limited runtime \citep{2020arXiv200307293S}. In this paper, we study simultaneously the acceleration of electrons, protons and helium ions at a nonrelativistic shock via a PIC simulation in one spatial dimension for a long runtime to make the DSA developed. The distribution of the amplified magnetic field and the spectra of the accelerated particles from the simulation are presented. In Section \ref{sect:mechod}, the detail of the numerical model is given. The numerical results are indicated in Section \ref{sect:result}. Finally, the summary and the conclusion are presented in Section \ref{sect:discon}.

\section{Numerics}
\label{sect:mechod}

\begin{figure*}
        \centering
        \includegraphics[width=0.42\textwidth, height=0.25\textheight]{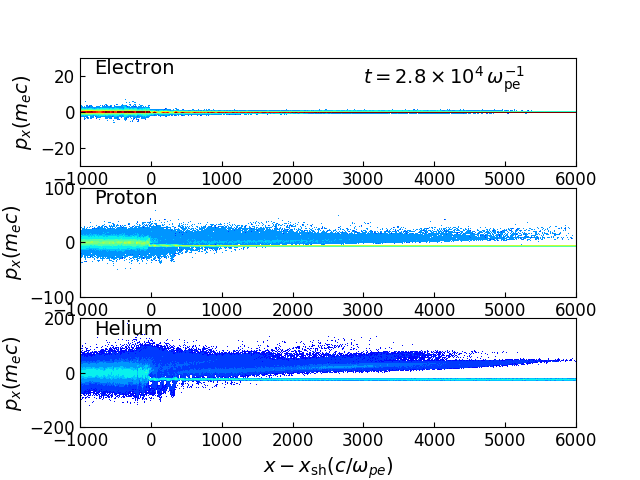}
        \includegraphics[width=0.5\textwidth, height=0.25\textheight]{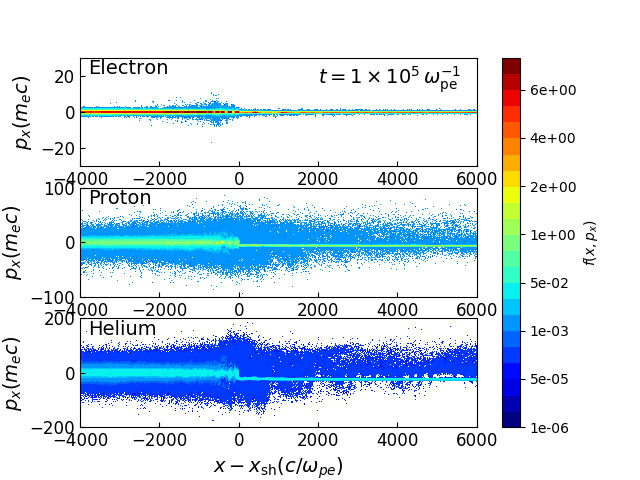}
        \caption{Phase space distributions of the electrons (top panels), the protons (middle panels), and the helium ions (bottom panels) for the shock at $t=2.8\times10^4\,\omega_{\mathrm{pe}}^{-1}$ (left panels), and $1.92\times10^5\,\omega_{\mathrm{pe}}^{-1}$ (right panels), respectively.}
        \label{fig:phase28}
\end{figure*}

\begin{figure*}
        \centering
        \includegraphics[width=0.42\textwidth, height=0.25\textheight]{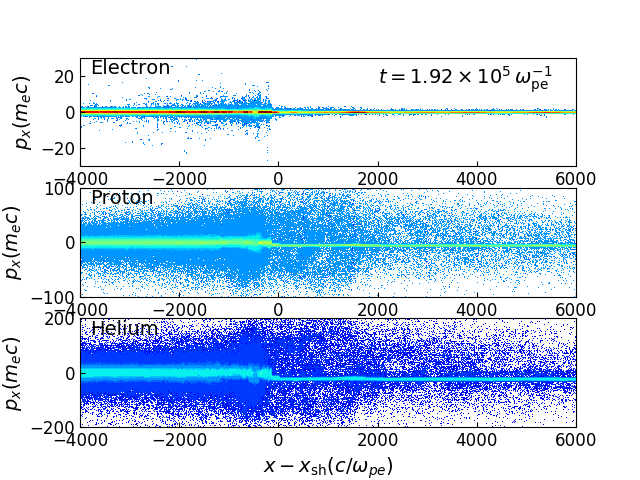}
        \includegraphics[width=0.5\textwidth, height=0.25\textheight]{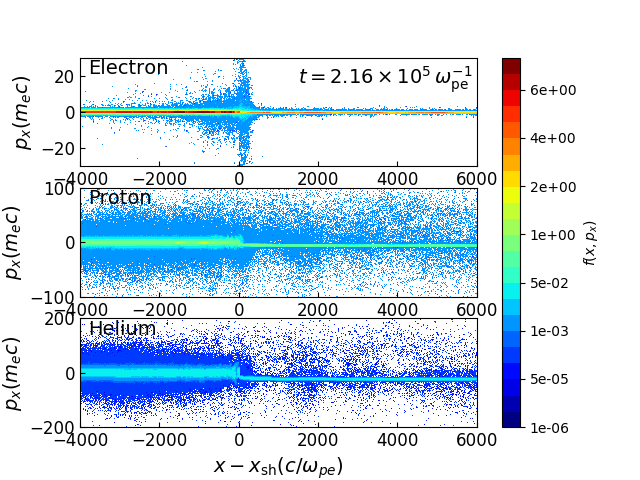}
        \caption{Phase space distributions of the electrons (top panels), the protons (middle panels), and the helium ions (bottom panels) for the shock at $t=1.92\times10^5\,\omega_{\mathrm{pe}}^{-1}$ (left panels), and $2.16\times10^5\,\omega_{\mathrm{pe}}^{-1}$ (right panels), respectively.}
        \label{fig:phase192}
\end{figure*}

Particle acceleration in a nonrelativistic quasiparallel shock is numerical investigated via PIC simulation using the code SMILEI \citep{2018CPC}, and the simulation is performed in 1D3V (one space dimension and three momentum dimensions) with reflection boundaries both for the particles and for the field. A reduced proton-to-electron mass ratio of $m_{\mathrm{p}}/m_{\mathrm{e}}=30$ and the helium mass is $m_{\mathrm{He}} = 4\,m_{\mathrm{p}}$ are adopted.  Initially, the plasma, which consists of protons ($A=1, Z=1$) with a number density of $n_{\mathrm{p}}=n_{\mathrm{0}}=0.1\,\mathrm{cm}^{-3}$ and a charge of $e$, helium ions (He$^{2+}$, $A=4, Z=2$) and electrons, has a velocity of $v_{\mathrm{p}}= 0.2 c$ flowing towards $-\hat{x}$, where $c$ is the speed of light. The density of the helium ions is $n_{\mathrm{He}}=0.1n_{\mathrm{0}}$, which is comparable to the helium-to-proton ratio for the cosmic rays in the Earth vicinity, and that of the electrons is $n_{\mathrm{e}}=1.2 n_{\mathrm{0}}$ to ensure electric neutrality of the plasma. The charges of the helium ions and the electrons are $2e$ and $-e$, respectively. The three components are firstly set to be in thermal equilibrium with a temperature of $T_0 = 2\times10^{-3}\,m_{\mathrm{e}}c^2/k_{\mathrm{B}}$, where $k_{\mathrm{B}}$ is the Boltzmann constant. $64$ particles per cell per species are used in the simulation. Fully kinetic PIC simulations must resolve the electron plasma frequency $\omega_{\mathrm{e}} = \sqrt{4 \pi n_{\mathrm{e}}e^2/m_{\mathrm{e}}} = \sqrt{1.2}\omega_{\mathrm{pe}}$, where $\omega_{\mathrm{pe}}^{-1} = \sqrt{m_{\mathrm{e}}/4 \pi n_{\mathrm{0}}e^2}$  is the time unit in the simulation. In this paper, the spatial extent along $x$ is $6.4\times10^4\,c/\omega_{\mathrm{pe}}$ with a resolution of $\Delta x = 0.05 c/\omega_{\mathrm{pe}}$ for the field, and the time step is adopted to be $0.02 \, \omega_{\mathrm{pe}}^{-1}$. Moreover, the simulation uses a total of $\sim 2.5\times10^8$ particles with $64$ particles per cell per species.

Particle acceleration in a shock relates to the mach number, the obliquity and the magnetization of it, and we investigate the acceleration of the particles composed of the three species at a nonrelativistic quasiparallel shock with a median sonic/Alfv\'{e}nic Mach number. With a compression radio of $r=4$ for the shock, the velocity of the upstream plasma in the shock rest frame is $v_{\mathrm{sh}}=v_{\mathrm{p}} r/(r-1) = 0.27 c$.  With a density of $n_{\mathrm{p}}$ and a temperature of $T_0$ for the protons, the sound speed is $v_{\mathrm{s}} = \sqrt{\gamma k_{\mathrm{B}} T_0/m_0} = 5.73\times 10^8\,\mathrm{cm\,s}^{-1}$ for the adiabatic index $\gamma=5/3$, where $m_0=(n_{\mathrm{p}} + 4 n_{\mathrm{p}})m_{\mathrm{p}}/(n_{\mathrm{e}} + n_{\mathrm{p}} + n_{\mathrm{He}}) $ is the averaged molecular mass, and the sonic Mach number is $M_{\mathrm{s}} = v_{\mathrm{sh}}/v_{\mathrm{s}} = 14$.  The initial magnetic field $\bf{B} = B_0(\cos \theta \hat{x} + \sin \theta \hat{y} )$ has a strength of $B_0=90\,\mathrm{G}$ with an inclination angle of $\theta = 30^\circ$, and the Alfv\'{e}n Mach number is $M_{\mathrm{A}}  = v_{\mathrm{sh}}/v_{\mathrm{A}} = 19.5$, where $v_{\mathrm{A}}$ is the Alfv\'{e}n speed.

\section{Results}

\label{sect:result}

\begin{figure}
        \centering
        \includegraphics[width=0.45\textwidth]{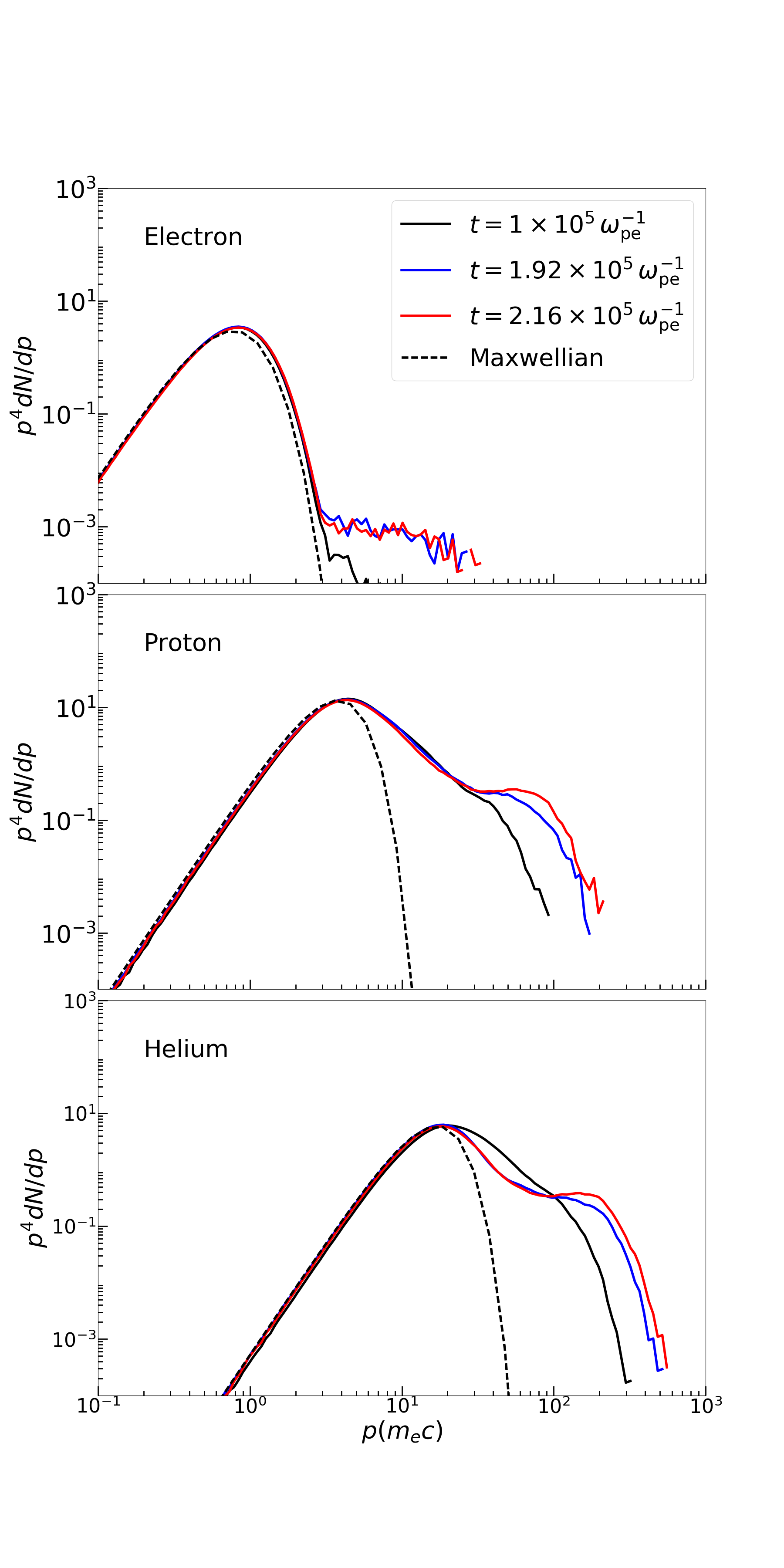}
        \caption{Electron (top panel), proton (middle panel) and helium ion  (bottom panel)  spectra for the particles in the downstream region $2200 -  4200\,c/\omega_{\mathrm{pe}}$ to the shock  at $t=1\times10^5\,\omega_{\mathrm{pe}}^{-1}$, $1.92\times10^5\,\omega_{\mathrm{pe}}^{-1}$  and $2.16\times10^5\,\omega_{\mathrm{pe}}^{-1}$, respectively. The dashed lines correspond to relativistic Maxwellian distributions with a temperature of $0.12\,m_{\mathrm{e}} c^2/k_{\mathrm{B}}$ both for the electrons and for the protons, and $0.6\,m_{\mathrm{e}} c^2/k_{\mathrm{B}}$ for the helium ions. }
        \label{fig:particle28}
\end{figure}

\begin{figure}
        \centering
        \includegraphics[width=0.45\textwidth]{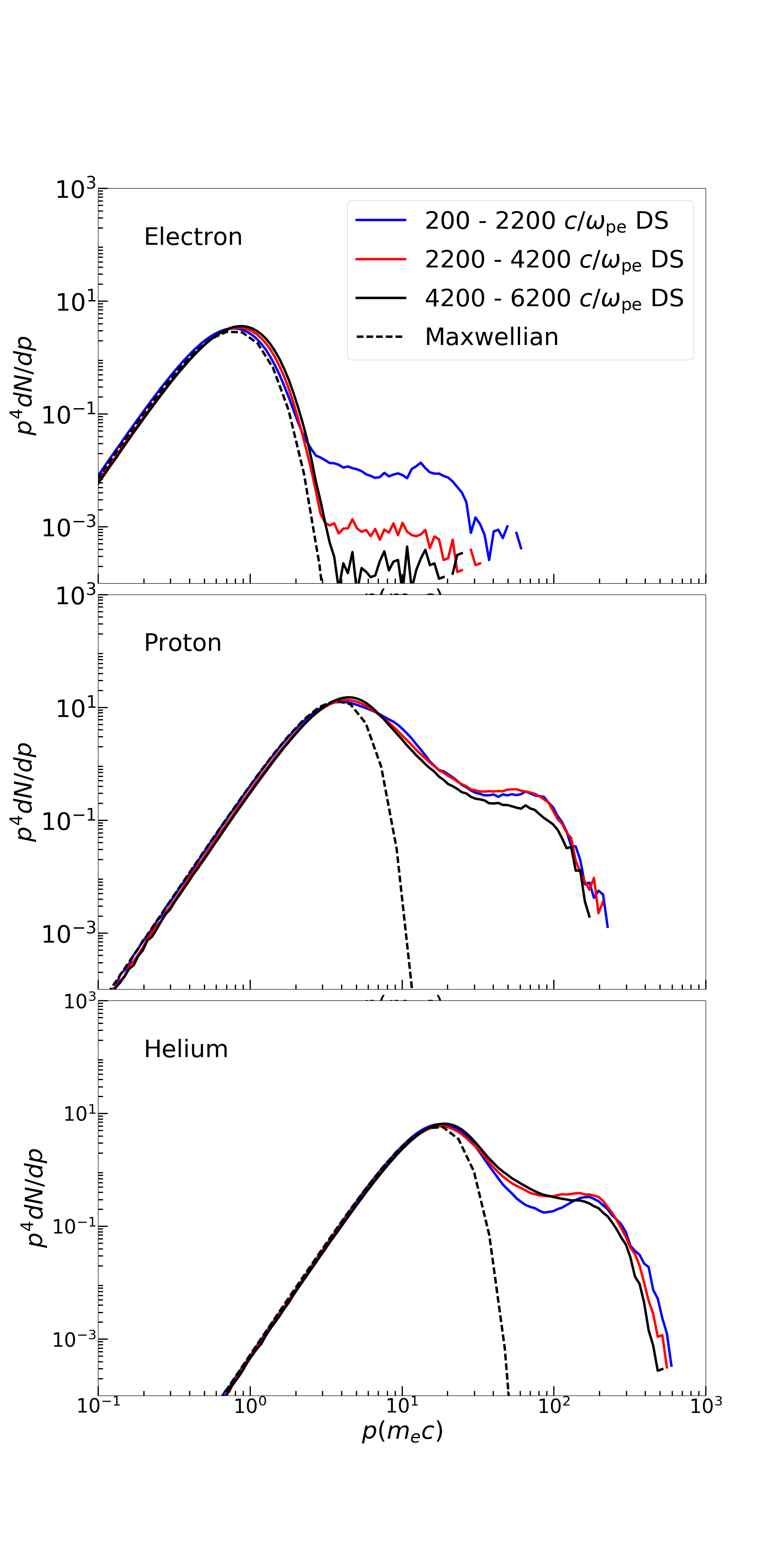}
        \caption{Electron (top panel), proton (middle panel) and helium ion  (bottom panel)  spectra for the particles  in the downstream regions $200 -  2200\,c/\omega_{\mathrm{pe}}$, $2200 -  4200\,c/\omega_{\mathrm{pe}}$ and $4200 -  6200\,c/\omega_{\mathrm{pe}}$ to the shock at $t=2.16\times10^5\,\omega_{\mathrm{pe}}^{-1}$, respectively. The temperatures of dashed lines representing the relativistic Maxwellian distributions are the same as Figure \ref{fig:particle28}. }
        \label{fig:fardown}
\end{figure}

A shock is generated since the reflected particles at the left boundary ($x=0$) propagate into the incoming plasma towards $-\hat{x}$ direction. Figure \ref{fig:shockstruc} shows the distributions of the proton number density, the self-generated transverse magnetic field $\delta B_y = B_y - B_0\sin \theta$ and $\delta B_z = B_z$ upstream and downstream of the shock at $t=1\times10^5\,\omega_{\mathrm{pe}}^{-1}$, $1.92\times10^5\,\omega_{\mathrm{pe}}^{-1}$ and $2.16\times10^5\,\omega_{\mathrm{pe}}^{-1}$, respectively. The particles flowing into the shock are heated and compressed, and a compression ratio of $\sim 4$ is obtained farther downstream of shock. The location of the shock is $x_{\mathrm s} = 6.25\times 10^3\,c/\omega_{\mathrm{pe}}$ and $\sim 1.28\times 10^4\,c/\omega_{\mathrm{pe}}$ at a time of $1\times10^5\,\omega_{\mathrm{pe}}^{-1}$ and $2.16\times10^5\,\omega_{\mathrm{pe}}^{-1}$, respectively, and it transports towards $+x$ with a velocity of $\sim 0.06\,c$. As indicated at the end of the simulation, instabilities are effectively generated around the shock, and a precursor, in which waves are effectively emitted, ahead of the shock is produced.

Figure \ref{fig:phase28} indicates the $x - p_x$ phase space distributions for the  electrons, the protons and the helium ions at the three times same as Figure \ref{fig:shockstruc}, where $f(x, p_x)$ is the number density of particles in the phase space of $p_x - x$. Since the three particle species possess different inertia entering the shock, an ambipolar electric field is induced around the shock \citep{2014ApJ...794..153G}. A part of the protons and the helium ions do not have enough kinetic energy to overcome the potential energy to cross the shock, and they are reflected by the shock. As a result, a stream of particles with positive $p_x$ upstream of the shock is prominent in the phase space distributions for both the protons and the helium ions. At $t=2.8\times10^4\,\omega_{\mathrm{pe}}^{-1}$, the charged particles interact with the self-generated magnetic field in the precursor region, which results in a part of them having negative $p_x$ as indicated in  Figure \ref{fig:phase28}. At later times indicated in Figure \ref{fig:phase192}, the turbulent magnetic field upstream of the shock is amplified more prominently, and the interaction of the particles with the waves becomes more effectively to make more particles to possess negative $p_x$.

All of the three particle species can be accelerated in the shock during interacting with the self-generated magnetic field. Figure \ref{fig:particle28} illustrates the particle spectra downstream of the shock ($- 4200\,c/\omega_{\mathrm{pe}} < x-x_{\mathrm {sh}} < - 2200\,c/\omega_{\mathrm{pe}} $ to the shock) at the three times. In the region, all of the three species develop a non-thermal component attach to the Maxwellian distribution, and the non-thermal spectra  become harder with time. The maximum energies of the accelerated electrons, protons and helium ions increase with time, and they are limited by the simulation time. The electrons and the protons downstream of the shock are heated to a similar temperature of $ T_{\mathrm{e}}=T_{\mathrm{p}} \sim 0.12\,m_e c^2/k_{\mathrm{B}}$, which implies thermal equilibration is effectively established between the two species \citep{2013ApJ...765..147P}. Alternatively, since the helium ions have more kinetic energy to convert into thermal energy, the downstream temperature for them is $\sim 0.6\,m_{\mathrm{e}} c^2/k_{\mathrm{B}}$, which is roughly consistent with the relation, i.e., $T_{\mathrm{i}}=A_{\mathrm{i}}T_{\mathrm{p}}$, where $T_{\mathrm{i}}$ and $A_{\mathrm{i}}$ are the downstream temperature and the atomic mass of ion specie \citep{2017PhRvL.119q1101C}.  Based on the Rankine-Hugoniot relation for a steady, plane-parallel shock without magnetic field with a Mach number of $M_1$, the post-shock temperature is
\begin{equation}
T_{2} = \frac{\left [(\gamma-1)M_1^2 + 2 \right ]\left[ 2\gamma M_1^2 - (\gamma-1)\right ]}{(\gamma + 1)^2 M_1^2}T_1,
\label{eq:tdown}
\end{equation}
where $T_1$ is the pre-shock temperature. The post-shock temperature from Eq.\ref{eq:tdown} is $0.124\, m_{e}c^2/k_{\mathrm{B}}$ for the shock with $T_1=T_0$, $\gamma=5/3$ and $M_1=M_s=14$, which is consistent with that derived from the simulation for the protons and the electrons. 

For the electrons, the non-thermal distribution begins at an momentum of $ p_{\mathrm{inj}}\sim 3\,m_e c$, and a power-law distribution with an index of $\sim4$ in the momentum space for the particles in the downstream region $2200 -  4200\,c/\omega_{\mathrm{pe}}$ to the shock, i.e., $dN/dp \propto p^{-4}$, is developed between $4\,m_e c $ and $ 20\,m_e c$ at $t = 2.16\times10^5\,\omega_{\mathrm{pe}}^{-1}$. The spectra of the protons and the helium ions are also harder at later times. At $t = 2.16\times10^5\,\omega_{\mathrm{pe}}^{-1}$, a power-law distribution with an index of $\sim4$ with momentum from $30\,m_e c$ to $\sim 100\,m_e c$ both for the protons and for the helium ions in the momentum range $100 -  200 \,m_e c$ in the downstream region.  For comparison, the spectra of the protons and the ions downstream of the shock in \citet{2020arXiv200307293S} thermal plus a supra-thermal tail, and  power law distribution in the particle spectra are not formed due to the limited runtime.

Figure \ref{fig:fardown} shows the spectra of the electrons (left panel), the protons (middle panel) and the helium ions (right panel) in the downstream regions $200 -  2200\,c/\omega_{\mathrm{pe}}$, $2200 -  4200\,c/\omega_{\mathrm{pe}}$ and $4200 -  6200\,c/\omega_{\mathrm{pe}}$ to the shock at $t=2.16\times10^5\,\omega_{\mathrm{pe}}^{-1}$, respectively. All the three species of the particles in the plasma develop a non-thermal spectrum with attach to the thermal distributions. As illustrated in Fig.\ref{fig:phase28}, most energetic electrons are effectively confined by the amplified magnetic field near the shock, so more  energetic electrons contributing to the non-thermal tail in the spectrum in closer region to the shock. Different with the electrons, the protons and the helium ions have much larger inertia, and the distributions of these two species are more homogeneous in the three downstream regions than the electrons. In Figure \ref{fig:fardown}, the spectra of the protons and the helium ions roll over at energies $E_{\mathrm max} \sim 0.9\times 10^2 m_e c$ and $\sim2\times 10^2 m_e c$, respectively, which is generally consistent with $E_{\mathrm max} \propto Z$.

The acceleration efficiency of the shock can be represented by the ratio of the kinetic energy contained in the nonthermal particles in a region to the bulk flow energy of the plasma before encountering the shock, i.e.,
\begin{equation}
\eta = \frac{\int_{p_{\mathrm{inj}}}E_{\mathrm{k}}(p) \frac{dN}{dp} dp}{0.5 (\tilde{n}_{\mathrm{e}} m_{\mathrm{e}} + \tilde{n}_{\mathrm{p}} m_{\mathrm{p}} + \tilde{n}_{\mathrm{He}} m_{\mathrm{He}})  v_{\mathrm{p}}^2},
\label{eq:effi}
\end{equation}
where $ E_{\mathrm{k}}(p)$ is the kinetic energy of the particle with a momentum of $p$, $\tilde{n}_{\mathrm{e, p, He}}$ is the averaged number density of the electrons, the protons, and the helium ions in the region, respectively. At $t=2.16\times10^5\,\omega_{\mathrm{pe}}^{-1}$, the acceleration efficiencies $\eta$ for the electrons, the protons and the Helium ions in the downstream region $200 -  2200\,c/\omega_{\mathrm{pe}}$ to the shock are $0.5\%$, $7.2\%$ and $5.4\%$, respectively, and about $13.1\%$ of the upstream bulk energy can be transferred into the nonthermal component of the particles in this region.

Charged particles can be accelerated at nonrelativistic quasiparallel shocks due to the joint effect of shock drift acceleration (SDA) and DSA \citep{2015PhRvL.114h5003P}. The magnetic gradient at the shock causes the particles drift along the shock surface, and they can be accelerated by the motional electric field via SDA \citep{2001PASA...18..361B,2006A&A...454..969M,2013ApJ...765..147P,2014ApJ...794..153G}. When a particle is injected into DSA, it gains energy as interacting with the upstream magnetic fluctuations. In a nonrelativistic quasiparallel shock, protons can be injected into DSA after preheated via SDA for several gyrocycles \citep{2015PhRvL.114h5003P}. Alternatively, electrons, which are first accelerated via SDA, have a hybrid acceleration process in which they  can be reflected by the upstream waves near the shock to encounter more processes of SDA before finally injected into DSA \citep{2015PhRvL.114h5003P}. As illustrated in Figure \ref{fig:phase28}, at $t=2.16\times10^5\,\omega_{\mathrm{pe}}^{-1}$, a significantly fraction of the protons and the helium ions farther upstream of the shock have been scattered by the upstream irregularities to possess a negative $p_x$, which indicates that they have been injected into DSA. However, the electrons are still in the hybrid stage up to $t=2.16\times10^5\,\omega_{\mathrm{pe}}^{-1}$, because most of the energetic electrons upstream of shock are confined within a region with an extension of just $\sim 200\,c/\omega_{\mathrm{pe}}$.

\section{summary and discussion}
\label{sect:discon}

We have performed a kinetic simulation on the acceleration of electrons, protons and helium ions (He$^{2+}$) in a  nonrelativistic collisionless quasiparallel shock with the helium-to-proton ratio comparable to the interstellar medium (ISM) in the Galaxy. The results show that all the three species downstream of the shock can be efficiently accelerated, and the spectra of all the three species in the farther downstream region have a power-law component with an index of $\sim 4$ in the momentum space attach to the thermal distribution, which is consistent with the expected value in the theory of DSA.  Due to the limited runtime, the downstream electrons near the shock have a power law spectrum from $\sim 3 m_{\mathrm e}c$ to $\sim 20 m_{\mathrm e}c$ at the end of the simulation, and the maximum energy of these accelerated electrons is lower than that in \citet{2015PhRvL.114h5003P}. Using PIC simulations, \citet{2020arXiv200307293S} studied shock formation and particle acceleration in the shock formed in a plasma composed of electrons, protons, helium and carbon
ions, and the property that the temperatures of the different ions downstream of the shock scale with the ratio of mass to charge is reproduced. This character of the downstream temperature for  protons and  helium ions is roughly reproduced at the end of the simulation in this paper. 

In this paper, at the end of the simulation with $t=2.16\times10^5\,\omega_{\mathrm{pe}}^{-1}$, for the shock with $M_{\mathrm{s}}= 14$,  $M_{\mathrm{A}} = 19.5$, and $m_p/m_e=30$, both the protons and the helium ions have entered the process of DSA, whereas the electrons are still undergo the hybrid process in which they are accelerated by SDA and DSA; moreover, the roll-over energies in the spectra of proton and He$^{2+}$ are generally proportional to the charge number $Z$, which means DSA is a rigidity-dependent mechanism to accelerate charged particles. Hybrid PIC simulations on the acceleration of different species of ions in a shock indicated that incompletely ionized heavy ions can be preferentially accelerated \citep{2017PhRvL.119q1101C}, which shed light on the heavy-ion enhancements in Galactic cosmic rays \citep{2011APh....34..447C}.

\section*{Acknowledgements}
We thank the anonymous referee for the comments. JF is supported by National Natural Science
Foundation of China (NSFC) under grants  11873042, U2031107, the Program of Yunnan University (WX069051, 2017YDYQ01), the grant from Yunnan Province (YNWR-QNBJ-2018-049)
and the National Key R\&D Program of China under grant No.2018YFA0404204.
HY is supported by the NSFC through grant 12063004.

\section*{Data availability}

The data produced in this paper will be shared on reasonable request to the corresponding author.

\setlength{\bibhang}{2.0em}
\setlength\labelwidth{0.0em}
\bibliography{heaccmn.bbl}

\begin{thebibliography}{}
\providecommand{\href}[2]{#2}
  \providecommand{\doi}[1]{\href{http://dx.doi.org/#1}{doi:#1}}
  \providecommand{\eprint}[1]{\href{http://arxiv.org/abs/#1}{arXiv:#1}}

\bibitem[\protect\citeauthoryear{{Acero} et~al.,}{{Acero}
  et~al.}{2010}]{2010A&A...516A..62A}
Acero, F. et al. 2010, \aap, 516, 62


\bibitem[\protect\citeauthoryear{{Ball \& Melrose}}{{Ball \& Melrose}}{2001}]{2001PASA...18..361B}
Ball, L., \& Melrose, D. B., 2001, PASA, 18, 361.

\bibitem[\protect\citeauthoryear{{Bamba} et~al.,}{{Bamba}
  et~al.}{2003}]{2003ApJ...589..827B}
Bamba, A., Yamazaki, R., Ueno, Masaru., \& Koyama, K. 2003, \apj, 589, 827

\bibitem[\protect\citeauthoryear{{Bell},}{{Bell}}{1978}]{1978MNRAS.182..147B}
Bell, A. R. 1878, \mnras, 182, 147



\bibitem[\protect\citeauthoryear{{Blandford \& Ostriker}}{{Blandford \& Ostriker}}{1978}]{1978ApJ...221L..29B}
Blandford, R. D., \& Ostriker, J. P. 1978, \apj, 221, L29

\bibitem[\protect\citeauthoryear{{Caprioli} et~al.,}{{Caprioli}
  et~al.}{2011}]{2011APh....34..447C}
Caprioli, D., Blasi, P., \& Amato, E. 2011, , Astropart. Phys., 34, 447

\bibitem[\protect\citeauthoryear{{Caprioli \& Spitkovsky}}{{Caprioli \& Spitkovsky}}{2014a}]{2014ApJ...783...91C}
Caprioli, D., \& Spitkovsky, A., 2014a, \apj, 783, 91

\bibitem[\protect\citeauthoryear{{Caprioli \& Spitkovsky}}{{Caprioli \& Spitkovsky}}{2014b}]{2014ApJ...794...46C}
Caprioli, D., \& Spitkovsky, A. 2014b, \apj, 794, 46

\bibitem[\protect\citeauthoryear{{Caprioli} et~al.,}{{Caprioli}
  et~al.}{2017}]{2017PhRvL.119q1101C}
Caprioli, D., Yi, D. T., Spitkovsky, A. 2017, \prl, 119, 171101


\bibitem[\protect\citeauthoryear{{Derouillat} et~al.,}{{Derouillat}
  et~al.}{2018}]{2018CPC}
Derouillat, J., Beck, A., P\'{e}rez,F., et al.
2018, Comput. Phys. Commun. 222, 351.

\bibitem[\protect\citeauthoryear{{Guo} et~al.,}{{Guo}
  et~al.}{2014}]{2014ApJ...794..153G}
Guo, X., Sironi, L., Narayan, R. 2014, \apj, 794, 153

\bibitem[\protect\citeauthoryear{{Ha} et~al.,}{{Ha}
  et~al.}{2018}]{2018ApJ...864..105H}
Ha, J.-H., Ryu, D., Kang, H. et al. 2018, \apj, 864, 105

\bibitem[\protect\citeauthoryear{{Kumar \& Reville}}{{Kumar \& Reville}}{2021}]{2021ApJ...921L..14K}
Kumar, N., \& Reville, B., 2021, \apjl, 921, 14

\bibitem[\protect\citeauthoryear{{Long} et~al.,}{{Long}
  et~al.}{2003}]{2003ApJ...586.1162L}
Long, K. S., Reynolds, S. P., Raymond, J. C. et al. 2003, \apj, 586, 1162

\bibitem[\protect\citeauthoryear{{Mann} et~al.,}{{Mann}
  et~al.}{2006}]{2006A&A...454..969M}
Mann, G., Aurass, H., Warmuth, A. 2006, \aap, 454, 969

\bibitem[\protect\citeauthoryear{{Park} et~al.,}{{Park}
  et~al.}{2013}]{2013ApJ...765..147P}
Park, J., Ren, C., Workman, J. C., \& Blackman, E. G. 2013, \apj, 765, 147

\bibitem[\protect\citeauthoryear{{Park} et~al.,}{{Park}
  et~al.}{2015}]{2015PhRvL.114h5003P}
Park, J., Caprioli, D., Spitkovsky, A. 2015, \prl, 114, 085003

\bibitem[\protect\citeauthoryear{{Reynoso} et~al.,}{{Reynoso}
  et~al.}{2013}]{2013AJ....145..104R}
Reynoso, E. M., Hughes, J. P., \& Moffett, D. A. 2013, \aj, 145, 104

\bibitem[\protect\citeauthoryear{{Rothenflug} et~al.,}{{Rothenflug}
  et~al.}{2004}]{2004A&A...425..121R}
Rothenflug, R.; Ballet, J.; Dubner, G. et al. \aap, 2004, 425, 121

\bibitem[\protect\citeauthoryear{{Schreiner} et~al.,}{{Schreiner}
  et~al.}{2020}]{2020arXiv200307293S}
Schreiner, C., Kilian, P., Spanier, F. et al. 2020, arXiv:200307293

\bibitem[\protect\citeauthoryear{{Xu} et~al.,}{{Xu}
  et~al.}{2020}]{2020ApJ...897L..41X}
Xu, R., Spitkovsky, A., Caprioli, D. 2020, \apjl, 897, 41


\bibitem[\protect\citeauthoryear{{Vink},}{{Vink}}{2012}]{2012A&ARv..20...49V}
Vink, J. 2012, A\&ARv, 20, 49


\end{thebibliography}

\end{document}